\documentclass[aps,prb,twocolumn,showpacs,amsfonts,superscriptaddress]{revtex4-1}
\usepackage{amsmath}
\usepackage{amssymb}
\usepackage{dsfont}
\usepackage{graphicx}
\usepackage{multimedia}
\usepackage{color}
\usepackage{xcolor}
\usepackage{ bbold }
\usepackage{bm}
\usepackage{bbm}
\newcommand{\mb}{\mathbf}

\begin{document}

\title{Switchable Josephson current in junctions with spin-orbit coupling}

\author{B. Bujnowski}
\email{bogusz.bujnowski@gmail.com}
\affiliation{Donostia International Physics Center (DIPC) - Manuel de Lardizabal 5, E-20018 San Sebasti\'{a}n, Spain}
       
\author{R. Biele}
\affiliation{Max Bergmann Center of Biomaterials, TU Dresden, 01062 Dresden, Germany}
\affiliation{Institute for Materials Science, TU Dresden, 01062 Dresden, Germany}

\author{F. S. Bergeret}
\email{fs.bergeret@csic.es}
\affiliation{Donostia International Physics Center (DIPC) - Manuel de Lardizabal 5, E-20018 San Sebasti\'{a}n, Spain}
\affiliation{Centro de F\'isica de Materiales (CFM-MPC) Centro Mixto CSIC-UPV/EHU - 20018 Donostia-San Sebastian, Basque Country, Spain}

\date{\today}

\begin{abstract}  
We study  the Josephson current  in  two types of lateral junctions  with spin-orbit coupling and an exchange field. The first system (type 1 junction)  consists of superconductors with heavy metal interlayers linked by a ferromagnetic bridge, such that the spin-orbit coupling is finite only at the superconductor/heavy metal interface. In the second type (type 2) of system we assume that the spin orbit coupling is finite in the bridge region. The length of both junctions is larger than the magnetic decay length such that the Josephson current is carried uniquely by the long-range  triplet component of the condensate.   The latter is generated by the spin-orbit coupling via two mechanisms, spin precession and inhomogeneous spin-relaxation. 
We show that the current can be controlled  by rotating the magnetization of the bridge  or by tuning the strength of the spin-orbit coupling in type 2 junctions., and 
also discuss how the ground-state of the junction can be tuned  from a $0$ to a $\pi$ phase difference between the superconducting electrodes. 
In  leading order in the spin-orbit coupling, the spin precession dominates the behavior of the triplet component and both 
 junctions behave similarly.  However, when spin relaxation effects are included junction of type  2 offers a wider parameter range in which  $0$-$\pi$ transitions take place. 
\end{abstract}

\maketitle
\section{Introduction}
The interplay between superconductivity and  ferromagnetism  leads to triplet superconducting correlations \cite{Bergeret2001a,Bergeret2005,Eschrig2011,Linder2015}. The simplest setup for the generation of a triplet component is a superconductor (S)-ferromagnet (F) heterostructure with a homogeneous exchange field. The superconducting singlet Copper pairs can   penetrate the ferromagnet,   and due to the local   exchange field, are  partially converted   into triplet pairs  with the total spin projection zero with respect to the local exchange field. Oscillations of the triplet correlations in the F region lead to the well understood effect of the sign reversal of the critical current,  the so-called  $0-\pi$ transition\cite{Bu1977,Panyukov1982,Ryazanov2001,Volkov2001,Kontos2002}.  In a diffusive monodomain F, both singlet and triplet  correlations decay on the magnetic length scale $\xi_h=\sqrt{D/h}$, where $h$ is the magnitude of the exchange field and $D$ is the diffusion constant. For conventional Ss and typical exchange field strengths, $\xi_h$ is much shorter than the thermal length scale of decay $\xi_\omega\approx\sqrt{D/T}$ in a non-magnetic system.   On the other hand, triplet components with  non-zero spin projection, are not affected by its pair breaking effect and would decay over  a length scale comparable to $\xi_\omega$. Such long-range triplet components (LRTC) can be generated due to inhomogeneities of the exchange field \cite{Bergeret2001a,Bergeret2005,Linder2015} or due to the presence of spin-orbit coupling (SOC) and a homogeneous exchange field \cite{Bergeret2013,Bergeret2014}.

The prediction of LRTC  in S/F hybrid structures has stimulated multiple experimental works \cite{Robinson2011,Anwar2010,Wang2014,Gingrich2012,Robinson2012,Chiodi2013,Pal2014,Robinson2014,Kalcheim2012,Banerjee2014,Khaire2010}.  More recently, transverse vertical heterostructures with in-plane magnetic fields and SOC materials have been experimentally explored but the long-range correlations due to SOC  have not been observed \cite{Satchell2018, Banerjee2018, Satchell2019}.  In accordance with  previous theoretical works \cite{Bergeret2013,Bergeret2014} in  vertical multilayered SFS junctions the condition for the generation of a LRTC is quite restrictive. More suitable  for the observation of LRTC induced by the SOC are lateral structures where currents have also a component flowing in the direction parallel to the hybrid interface \cite{Liu2014,Arjoranta2016,Eskilt2019}.

\begin{figure}
   \begin{center} 
   \includegraphics[width=1.0\columnwidth]{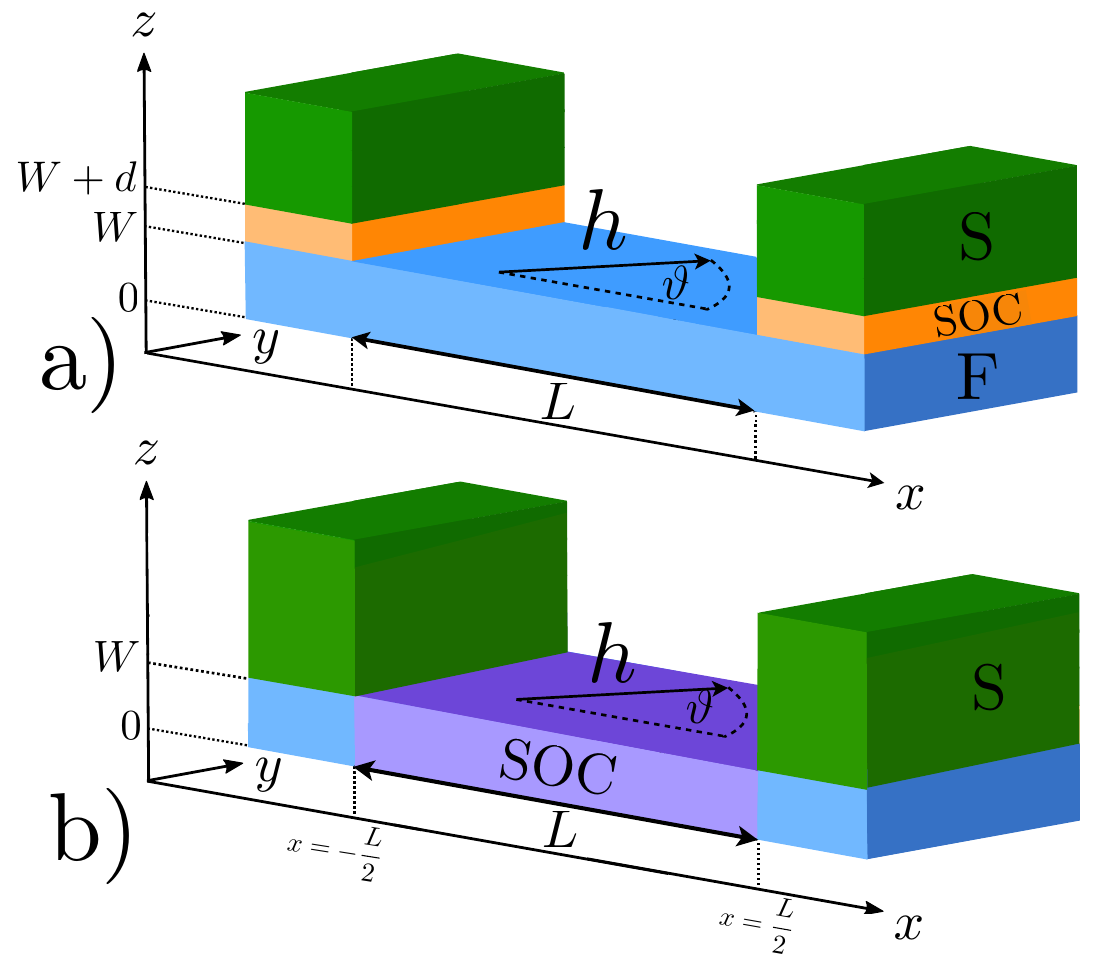}
   \caption{Schematic view of  the two junction types considered in the text. a) The junction of type 1 consists of two superconductors contacted to a  ferromagnet (F) via a material with strong SOC.
    The magnetization  of F, and hence the exchange field $\mb{h}$,  lies  in the  $x-y$ plane. b) For the junction of type 2 the  bridge   region connecting the two superconductors has a sizable SOC. 
   .}\label{junction}
   \end{center}
\end{figure}

In this work we present a study of the  Josephson current in lateral geometries with SOC of Rashba and Dresselhaus type and how to  control it via external fields in the diffusive regime. We focus on two types of junctions: One consists of  two superconducting electrodes on top of a ferromagnetic film, see Fig. \ref{junction}a). Between the two materials we assume there is an interlayer with a finite SOC.  Hereafter we refer to this junction as type 1 junction.  The junction of type 2, Fig. \ref{junction}b), consists of a similar lateral geometry, but the SOC is  finite  in the bridge region. Whereas type 1 junctions may correspond to junctions with a heavy metal interlayer, type 2 junctions describe, for example, a lateral Josephson junction made of  a 2D electron gas in the presence of a Zeeman field.
We assume that in both junctions the distance between the superconductor electrodes is larger than the magnetic decay length, such that the Josephson current is only carried by LRTC. The latter is generated  by the SOC via two mechanisms:  spin precession and inhomogeneous spin-relaxation and the current strongly depends  on the direction of the exchange or Zeeman fields. In addition, in type 2 junctions the Josephson current can also be tuned by a voltage gate that controls the strength of the Rashba SOC. 

We focus on the control of possible $0$-$\pi$ transitions. With the help of an analytical solution for type 1 junction in the case of small SOC, we first  show  that in leading order the LRTC is generated only by the spin-precession term and the junction remains in the $0$-state independently of the direction of the exchange field. The next leading order contribution to the current is  due to the inhomogeneous spin relaxation with a negative sign, such that for certain  directions of the exchange field the junction can switch to the $\pi$-state. In junctions of type 1 this only occurs if both, the Rashba and Dresselhaus  SOC are finite. 
In a second part we present numeric  calculations of the current for arbitrary SOC strength  that confirm these findings. 
In addition these calculations reveal that  type 2 junctions  allow for $0-\pi$ transitions in a wider range of SOC parameters. Specifically the transition can be induced by a pure Rashba or Dresselhaus SOC by changing their strengths.  This is a new 
possibility to induce $0$-$\pi$ transition by tuning the Rashba SOC strength, which is experimentally achievable by gating the SOC active material. Besides the interesting applications of such lateral junctions as  $0$-$\pi$ switchers,  they  can also be used  to detect the LRTC by measuring the changes of the  Josephson current as a function of the direction of the applied field or magnetization in a single junction. 

The work is organized  as follows:  In the next section \ref{section2} we present the basic equations describing diffusive Josephson junctions and we adapt these equations to the lateral junctions type 1 and 2. In section \ref{section3} we derive the analytical expression for the Josephson current in junction type 1 perturbatively, up to second order in the SOC parameter for semi-infinite leads. In section \ref{section4} we present numerical results for the Josephson current for both types of junctions  and compare them to the analytical results. Conclusions are given  in section \ref{section5}.


\section{Basic equations for diffusive Josephson junctions with SOC}\label{section2}
We consider  two spatially separated superconducting electrodes  on top of a non-superconducting material with either an  intrinsic exchange field, as in a ferromagnet,  or a Zeeman field induced by an external magnetic field.
We distinguish two different types of junctions:  one  with SOC  active layers just below the superconductors, Fig.\ref{junction}a),  that we refer to as junction type 1. The other junction with SOC  in the bridge region is referred to as junction type 2, and is shown in Fig.\ref{junction}b).

We assume that the proximity effect, {\it i.e.} the induced superconducting correlations in the bridge, is weak and that the system is in the diffusive regime. 
In this case spectral and transport properties of the junction can be accurately described by the 
linearized Usadel equation\cite{Usadel1970} generalized to linear in momentum SOC.  This equation  provides  the spatial dependence of the induced superconducting correlations in the non-superconducting region which is described  in terms of the anomalous Green's function $\hat{f}$\cite{Bergeret2013,Bergeret2014}\footnote{Eq. (\ref{Usadel_basic}) is written in the strict diffusive limit and do not take into account charge-spin conversion terms which are higher order in the momentum relaxation rate\cite{Bergeret2015,Konschelle2015}.}:
\begin{flalign}\label{Usadel_basic}
D\tilde{\nabla}_k^2\hat{f}+2|\omega_n|\hat{f}-i\text{sign}(\omega_n)\left\{\hat{h},\hat{f}\right\}=0.
\end{flalign}
Here $D$ is the diffusion constant, $\omega_n$ is the Matsubara frequency, $\tilde{\nabla}_k=\partial_k-i[\hat{\mathcal{A}}_k,\dots]$ is the covariant derivative with the SU(2) vector potential, $\hat{\mathcal{A}}_k=\frac{\hat{\sigma}^a}{2}\mathcal{A}_k^a$, describing the SOC and $\hat{h}=\hat{\sigma}^ah^a$ is the exchange field. Symbols with a $\hat{\phantom{.}}$ stand for operators in spin space and $\hat{\sigma}^a$ are the Pauli matrices. We use the Einstein summation convention and sum over repeated indices. The general form of the condensate  function in spin space is 
\begin{flalign}
 \hat{f}=f_s \hat{1}+f_t^a\hat{\sigma}^a
\end{flalign}
where $f_s$ is the singlet component and $f_t^a$ are the triplet components.  In our representation the short (long)-range triplet component corresponds to the component parallel (orthogonal) to the exchange or Zeeman field.

In order to describe hybrid interfaces between the superconductor and a substrate one needs boundary conditions for the Green's functions. 
We use here  the  Kupriyanov-Lukichev ones\cite{Kupriyanov1988} generalized for materials with SOC.
In its linearized form at an S/X interface they read\cite{Bergeret2013,Bergeret2014}:
\begin{flalign}\label{Kupriyanov}
\mathcal{N}_i \left[\tilde{\nabla}_i \hat{f}\right]_{S/X}=-\gamma f_{BCS}\hat{1},
\end{flalign}
where $X$ denotes  any  non-superconductor  material.
Here $\mathcal{N}_i$ is the $i$-th component of the interface normal, $\gamma$ is the interface transparency, $f_{BCS}=\frac{\Delta e^{i\varphi_i}}{\sqrt{\Delta^2-\omega_n^2}}$ is the anomalous Green's function in the bulk S region with the amplitude of the superconducting order parameter $\Delta$ and its phase $\varphi_i$  in the $i$-th electrode.  At the interface with  vacuum (V)  no current flows and  the boundary condition reads:
\begin{flalign}\label{vaccum}
\mathcal{N}_i \left[\tilde{\nabla}_i \hat{f}\right]_{X/V}=0\; .
\end{flalign}

Below  we determine  the Josephson current density in the bridge region for the two setups  depicted in Fig.\ref{junction}.  It can be expressed as\cite{Konschelle2015}:
 \begin{flalign}\label{current_full}
 	j=ie\pi N_0 D T \sum_{\omega_n}\text{Tr}\left\{\hat{f}\tilde{\nabla}\hat{\bar{f}}-\hat{\bar{f}}\tilde{\nabla}\hat{f}\right\}
 \end{flalign}
 where $N_0$ is the  density of states and $\hat{\bar{f}}=\hat{\sigma}^y\hat{f}^*\hat{\sigma}^y$.


\begin{figure*}[ht]
   \begin{center} 
      \includegraphics[width=1.9\columnwidth]{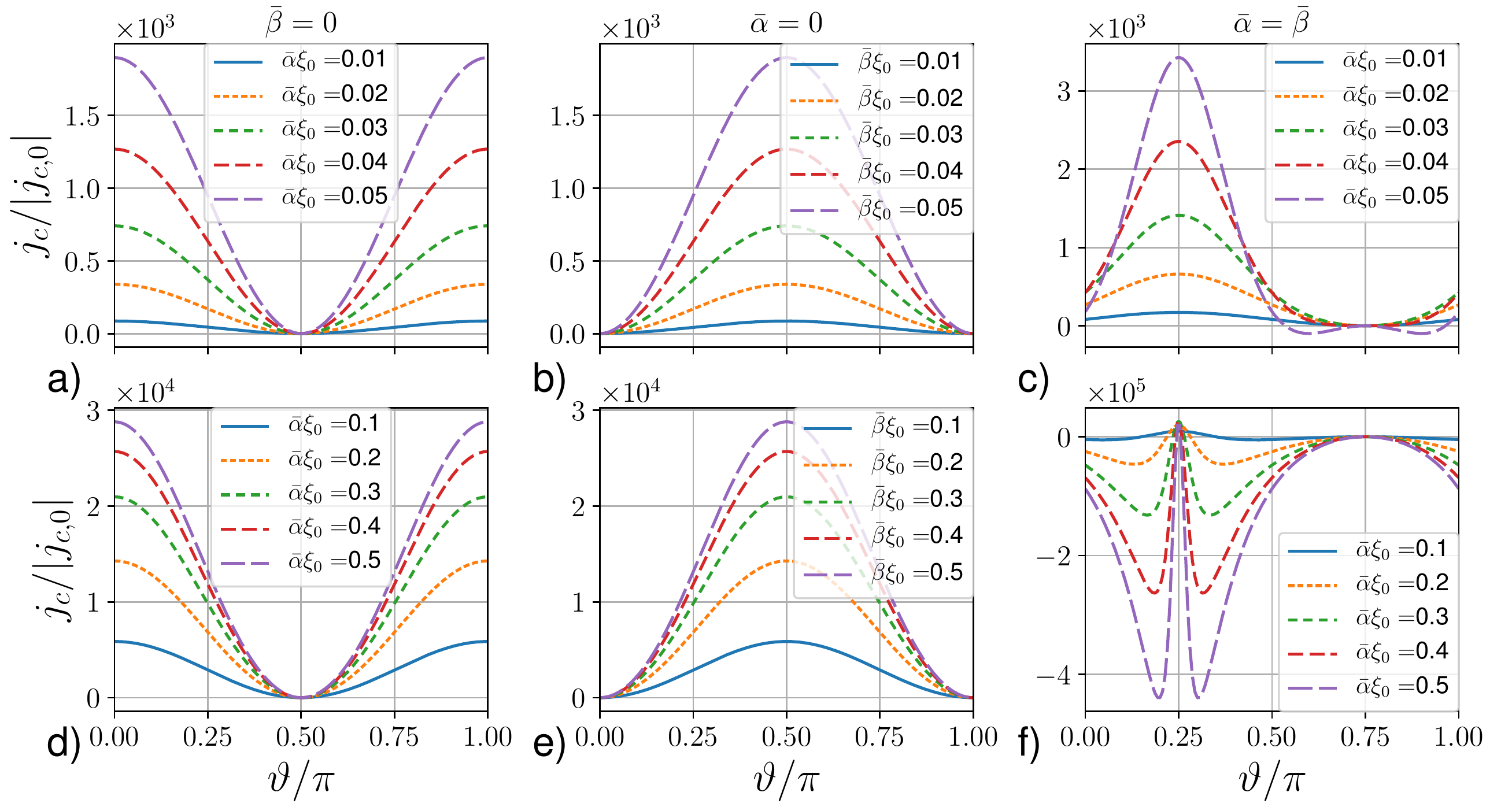}
      \caption{ Numerical results for the critical current as function of the orientation of an in-plane exchange field for junction of type 1.  Different curves correspond to different values of the  SOC parameters. In all plots we set $\bar{h}=10\Delta$, $L=5\xi_0$, $T=0.01 \Delta$ and the thickness of the SOC  and F layer are chosen such that  $d/W=1.$}\label{a_var1}
   \end{center}
\end{figure*}

We now re-write  Eqs.(\ref{Usadel_basic}) and (\ref{Kupriyanov}) for the specific case of junction of  type 1 and 2. Both  junctions are assumed to be  translational invariant in the $y$-direction. The order parameter is  a step-like-function  along  the $x$-direction,  with amplitude $\Delta$ at the S electrodes and zero at the bridge.  We denote with $\varphi$ the  phase difference   between the superconductors, such that 
\begin{flalign}\label{gap(x)}
 \Delta(x,z)=\Theta(z-(W+d))\Theta(|x|-L/2)\Delta e^{i\frac{\varphi}{2}\text{sign}(x)}.
\end{flalign}
The SOC fields are finite only in the SOC layers thus for the junction type 1 (Fig.\ref{junction}a)
\begin{flalign}\label{SOC(x)1}
 \mathcal{A}_k^a(x,z)=\Theta(W+d-z)\Theta(z-W)\Theta(|x|-L/2)\mathcal{A}_k^a
\end{flalign}
and for the junction type 2 with the SOC in the bridge region (Fig.\ref{junction}b) 
\begin{flalign}\label{SOC(x)2}
 \mathcal{A}_k^a(x,z)=\Theta(W-z)\Theta(z-W)\Theta(L/2-|x|)\mathcal{A}_k^a
\end{flalign}
with constant $\mathcal{A}_k^a$. We restrict ourselves to SOC of the Rashba and Dresselhaus type defined by the following vector potential:  $\hat{\mathcal{A}}_x=\beta/2\hat{\sigma}^x-\alpha/2\hat{\sigma}^y$ and $\hat{\mathcal{A}}_y=\alpha/2\hat{\sigma}^x-\beta/2\hat{\sigma}^y$.
Rashba SOC corresponds to terms proportional to $\alpha$ 
, while Dresselhaus SOC corresponds to terms proportional to $\beta$. For both junctions the exchange field has only finite components in the $x-y$ plane and is present in the region F,
\begin{flalign}\label{h(x)}
\hat{h}(x,z)=h(\cos{\vartheta}\hat{\sigma}^x+\sin\vartheta\hat{\sigma}^y)\Theta(z)\Theta(W-z),
\end{flalign}
where $h=\sqrt{h^a h^a}$.

To distinguish components that are parallel and perpendicular to the exchange field, {\it i.e.} short - and long-range components,  it is convenient to rotate Eqs.(\ref{Usadel_basic}) and (\ref{Kupriyanov}) by the unitary transformation $\mathcal{U}=e^{i\hat{\sigma}^x\frac{\vartheta}{2}}$. After the rotation the exchange field is fixed along the $x$-axis,
\begin{flalign}
 \mathcal{U}\hat{h}\mathcal{U}^\dagger=h\hat{\sigma}^x.
\end{flalign} 
Thus  in our notation the long-range triplet components  are those polarized in  $y$ and $z$ direction. Assuming for simplicity, that the thickness $d$ of the SOC interlayers (if present) and the bridge $W$ is small against the typical length on which $\hat{f}$ changes, we can integrate the Usadel equation along the $z$-direction\cite{Bergeret2014}. This reduces the initial two dimensional problem to an effective one dimensional one.

Here we illustrate  how the $z$-integration is carried out. Besides the  first term in Eq.(\ref{Usadel_basic}) , all other terms  do not contain a spatial derivative in the $z$-direction and therefore the integration results simply in the averaged value of $f$.  Integration of the first term of Eq.\ref{Usadel_basic} leads to 
 \begin{flalign}
	&\int_0^{W+d}\tilde{\nabla}_k^2\hat{f} dz=\int_0^{W+d}dz\left(\left(\partial_x^2+\partial_z^2\right)\hat{f}\right.\nonumber\\ &\left.-2i\left[\hat{A}_x,\partial_x\hat{f}\right]-\left[\hat{A}_k,\left[\hat{A}_k,\hat{f}\right]\right] \right)\approx\gamma f_{BCS}\nonumber\\
&+(W+d)\partial_x^2\hat{f} -2id\left[\hat{A}_x,\partial_x\hat{f}\right]-d\left[\hat{A}_k,\left[\hat{A}_k,\hat{f}\right]\right].
\end{flalign}
In the first step we have use the translational invariance in $y$-direction and our choice of  the SOC, which is step-wise constant.  In the second step we use the continuity of $\hat{f}$ in the $z$-direction and the boundary condition Eq.(\ref{Kupriyanov}) at the interface at $z=W+d$ as well as the boundary condition with the vacuum,  Eq.(\ref{vaccum}). The $z$-integration causes an averaging of the couplings that differs for the two junction types. We therefore present the final equations separately.
\subsection{ Usadel equations for type 1 lateral junction}
After performing the $z$-integration, and the rotation of Eq.(\ref{Usadel_basic}),  the resulting system of equations for the rotated anomalous Green's function $\hat{\tilde{f}}=\mathcal{U}\hat{f}\mathcal{U}^\dagger$ is 
\begin{flalign}\label{SOC1} 
 &D\left[\partial_x^2 \tilde{f}_s\right]-2|\omega_n|\tilde{f}_s-2i\text{sign}(\omega_n) \bar{h} \tilde{f}_t^x=\nonumber\\&-D\bar{\gamma} {f}_\text{BCS}e^{-i\text{sign}(x)\frac{\varphi}{2}},\\
 \label{eq:ftycomp}
&D\left[\partial_x^2 \tilde{f}_t^a+2{\bar{\mathcal{C}}}_x^{ab}\left(\partial_x \tilde{f}_t^b\right)\right]-2|\omega_n|\tilde{f}_t^a- D{\bar{\Gamma}}^{ab}\tilde{f}_t^b=\nonumber\\
&\delta_{x,a} 2i\text{sign}(\omega_n)\bar{h} \tilde{f}_s,
\end{flalign} 
where we introduced the Kronecker-Delta $\delta_{i,j}$, the components of the averaged spin precession tensor 
\begin{flalign}
\bar{\mathcal{C}}_k^{ab}=\varepsilon^{acb}\mathcal{A}_k^c d/(W+d)
\end{flalign}
and averaged Dyakonov-Perell (DP) spin relaxation tensor
\begin{flalign}
\bar{\Gamma}^{ab}=\left(\mathcal{A}_k^c\mathcal{A}_k^c \delta_{a,b}-\mathcal{A}_k^a\mathcal{A}_k^b\right) d/(W+d). 
\end{flalign}
The averaged coupling constants are defined as $\bar{h}^a=h^aW/(W+d)$, $\bar{\alpha}=\alpha d/(W+d)$, $ \bar{\beta}=\beta d/(W+d)$ and $\bar{\gamma}=\gamma/(W+d)$. The spatial dependence of the SOC fields, exchange field and order parameter in x-direction is not explicitly written, and is defined in  Eqs.(\ref{gap(x)}), (\ref{SOC(x)1}) and (\ref{h(x)}). In the rotated system the non vanishing spin precession tensor elements are
\begin{flalign}
	&{\bar{\mathcal{C}}}_{x}^{xz}=-{\bar{\mathcal{C}}}_{x}^{zx}=-\bar{\alpha}\cos(\vartheta)-\bar{\beta}\sin(\vartheta),\\ &{\bar{\mathcal{C}}}_{x}^{yz}=-{\bar{\mathcal{C}}}_{x}^{zy}=\bar{\alpha}\sin(\vartheta)-\bar{\beta}\cos(\vartheta).
\end{flalign}
The non zero elements of the DP spin relaxation tensor are 
\begin{flalign}
	&\bar{\Gamma}^{xx}(\vartheta)=\bar{\Gamma}^{yy}(-\vartheta)=\left(\bar{\alpha}^2+\bar{\beta}^2+\bar{\alpha}\bar{\beta}\sin(2\vartheta)\right)\frac{W+d}{d}\\
	&\bar{\Gamma}^{zz}(\vartheta)=\bar{\Gamma}^{xx}(\vartheta)+\bar{\Gamma}^{yy}(\vartheta),\\
	&\bar{\Gamma}^{xy}(\vartheta)=\bar{\Gamma}^{yx}(\vartheta)=2\bar{\alpha}\bar{\beta}\cos\left(2\vartheta\right)\frac{W+d}{d}
\end{flalign}

The solution of Eqs. (\ref{SOC1})-(\ref{eq:ftycomp}) and its covariant derivative are continuous at the boundaries $x=\pm L/2$ between the different regions thus:
 \begin{flalign}\label{SOC3}
  &\partial_x \tilde{f}_s \big|_{x=\pm\frac{L}{2}+0^-}=\partial_x \tilde{f}_s\big|_{x=\pm\frac{L}{2}+0^+},\\
  &\label{SOC5}\partial_x \tilde{f}_t^a\big|_{x=\pm\frac{L}{2}+0^\mp}=\left[\partial_x \tilde{f}_t^a +{\bar{\mathcal{C}}}_x^{ab} \tilde{f}_t^b\right]_{x=\pm \frac{L}{2}+0^\pm}.
 \end{flalign}
 The equations (\ref{SOC1}) and (\ref{eq:ftycomp}) together with the boundary conditions Eq.(\ref{SOC3}) and Eq.(\ref{SOC5}) fully determine the condensate within the limits of the mentioned approximations. 
 Finally the current in the bridge region is given by
  \begin{flalign}\label{eq:current_full1}
  	j&=4\pi eN_0DT\sum_{\omega_n}\text{Im}\left[\tilde{f}_s^*\partial_x \tilde{f}_s- (\tilde{f}_t^i)^*(\partial_x \tilde{f}_t^i)\right].
  \end{flalign}
\subsection{ Usadel equations for type 2 lateral junction}
For the junction type 2 the SOC coupling and the exchange field are finite over the whole bridge. Consequently $\bar{h}^a=h$, $\bar{\alpha}=\alpha$, $ \bar{\beta}=\beta$ and $\bar{\gamma}=\gamma/W$. Thus the $z$-integrated Usadel equation is like in 
Eqs.(\ref{SOC1}) and (\ref{eq:ftycomp}) where now 
\begin{flalign}
\bar{\mathcal{C}}_k^{ab}={\mathcal{C}}_k^{ab}=\varepsilon^{acb}\mathcal{A}_k^c 
\end{flalign}
and the DP spin relaxation tensor
\begin{flalign}
\bar{\Gamma}^{ab}={\Gamma}^{ab}=\mathcal{A}_k^c\mathcal{A}_k^c \delta_{a,b}-\mathcal{A}_k^a\mathcal{A}_k^b. 
\end{flalign}
The spatial dependence of the SOC fields is now  given by Eq.(\ref{SOC(x)2}). The solution of this system of equations is continuous and fulfills
\begin{flalign}\label{eq:bc_type2_01}
  &\partial_x \tilde{f}_s \big|_{x=\pm\frac{L}{2}+0^-}=\partial_x \tilde{f}_s\big|_{x=\pm\frac{L}{2}+0^+},\\
  &\label{eq:bc_type2_02}\left[\partial_x \tilde{f}_t^a +{{\mathcal{C}}}_x^{ab} \tilde{f}_t^b\right]_{x=\pm\frac{L}{2}+0^\mp}=\partial_x \tilde{f}_t^a\big|_{x=\pm \frac{L}{2}+0^\pm}.
 \end{flalign}
Thus for type 2 junctions, the condensate function is determined from Eqs.(\ref{SOC1})-(\ref{eq:ftycomp}) and Eqs.(\ref{eq:bc_type2_01})-(\ref{eq:bc_type2_02}). Finally  the  current through the junction is given by
 \begin{flalign}\label{eq:current_full2}
 	j&=4\pi eN_0DT\sum_{\omega_n}\text{Im}\left[\tilde{f}_s^*\partial_x \tilde{f}_s- (\tilde{f}_t^i)^*(\partial_x \tilde{f}_t^i)\right.\nonumber\\
	&\left.+(\tilde{f}_t^z)^*\left({\alpha} \tilde{f}_t^x+ {\beta} \tilde{f}_t^y\right)\right].
 \end{flalign}
\section{The Josephson current in type 1 Junctions: Analytical solution}\label{section3}
Her we focus on type 1 junctions in  the case when the exchange interaction is the dominant energy scale, $ D\alpha^2,D\beta^2, D\alpha\beta,T \ll h$. The junction  is larger than the magnetic length, $\xi_h$,  and hence  the current is solely determined by the LRTC, $\tilde{f}_t^y$ and $\tilde{f}_t^z$.  The other two components decay  over $\xi_h$ in the F region.

We  solve  Eqs.(\ref{SOC1}-\ref{eq:ftycomp})   perturbatively  up to second order in the  SOC fields, $\mathcal{A}_k^a$.  
In  zeroth order only  the singlet and triplet component parallel to the field,  $\tilde{f}_{t,0}^{x}$, are finite.  Their explicitly form is given in the appendix,  Eq.(\ref{eq:ftx_source}). 
In first order in the SOC  the component $\tilde{f}_t^z$  appears as a consequence of the precession term. Specifically  it  is determined by 
\begin{flalign}\label{eq:first_order}
\partial_x^2 \tilde{f}_{t,1}^{z}-\frac{2|\omega_n|}{D}\tilde{f}_{t,1}^{z}=-2\bar{\mathcal{C}}_x^{zx}\left(\partial_x \tilde{f}_{t,0}^{x}\right).
\end{flalign}
The component $\tilde{f}_t^y$  appears in second order of the SOC and satisfies:
\begin{flalign}\label{eq:second_order}
\partial_x^2 \tilde{f}_{t,2}^{y}-\frac{2|\omega_n|}{D}\tilde{f}_{t,2}^{y}=-2\bar{\mathcal{C}}_x^{yz}\left(\partial_x \tilde{f}_{t,1}^{z}\right)+ \bar{\Gamma}_{yx}(\vartheta)\tilde{f}_{t,0}^{x}.
\end{flalign}

\begin{figure*}
   \begin{center} 
      \includegraphics[width=1.9\columnwidth]{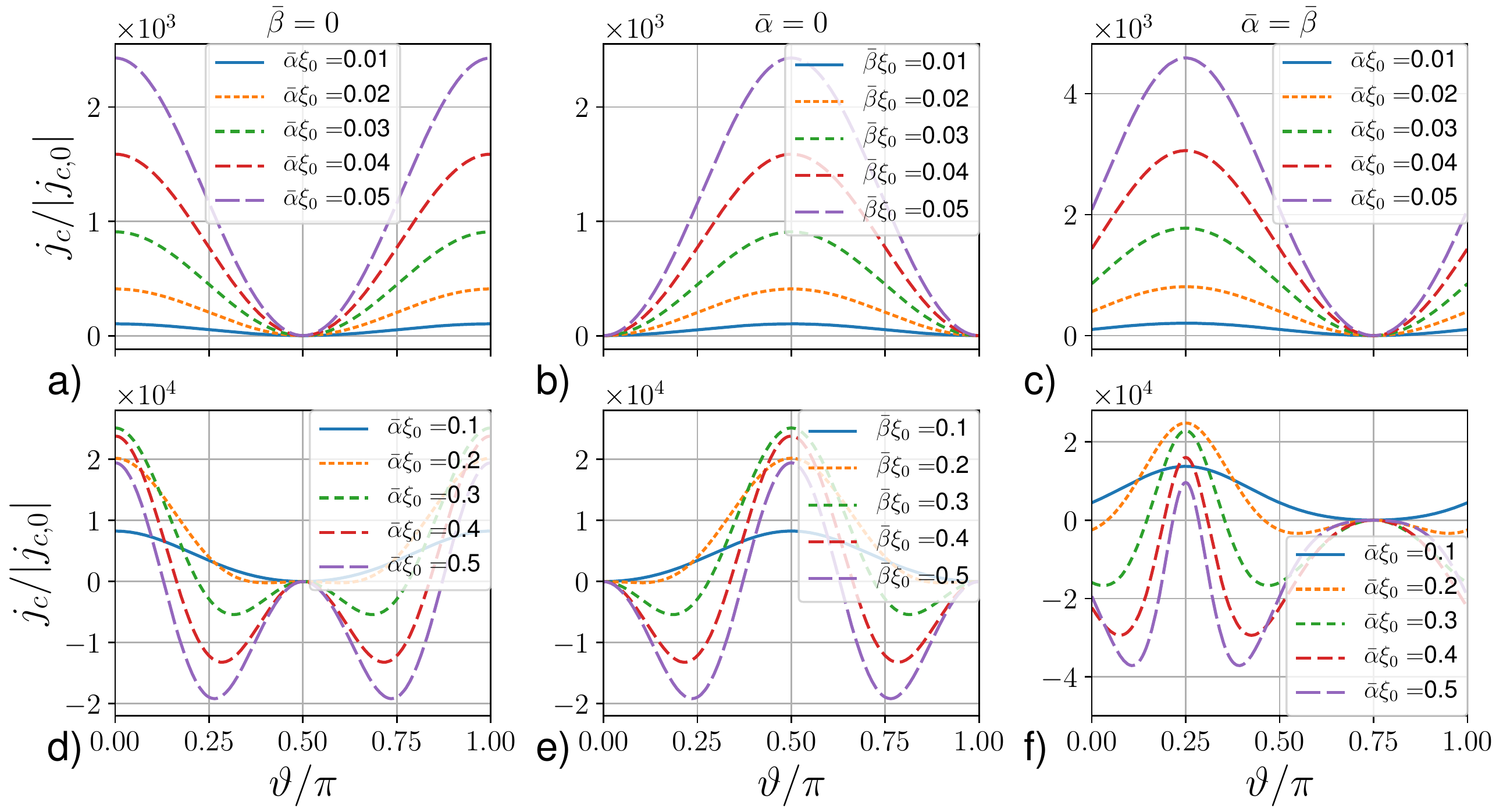}
	   \caption{ Numerical results for the critical current as function of the orientation of an in-plane exchange field for  in  junction of type 2.  Different curves correspond to different values of the  SOC parameters. In all plots we set $\bar{h}=10\Delta$, $L=5\xi_0$, $T=0.01 \Delta$.
  }\label{J2a_var2}
   \end{center}
\end{figure*}
The explicit expressions for these components  are given in the appendix, Eqs.(\ref{eq:first_order_sol}),
(\ref{eq:second_order_sol}). 
From these  solutions we obtain the current density in the F region.  The  current density Eq.(\ref{current_full}) is only due to the contribution of the long-range components $\tilde{f}_{t}^z$ and $\tilde{f}_{t}^y$ in Eq. (\ref{eq:current_full2}).   The maximum value of the Josephson current, {\it i.e.} the critical current $j_c$, 
is obtained at  $\varphi=\pi/2$:
\begin{flalign}\label{eq:current_precession}
	&j_c=j\left(\varphi=\frac{\pi}{2}\right)= \pi eN_0DT \sum_{\omega_n} {|f_x^b|^2}e^{-\kappa_\omega L}\times\nonumber\\
	&\left( \frac{\left(\bar{\alpha}\cos\vartheta+\bar{\beta}\sin\vartheta\right)^2}{2\kappa_\omega}- \frac{8\bar{\alpha}^2\bar{\beta}^2\cos^2 2\vartheta}{\kappa_\omega^3}\right)
\end{flalign}
where we define $\kappa_\omega=\sqrt{2|\omega_n|/D}$ and 
 \begin{flalign}
 f_{x}^b
   \approx -{i}\frac{{\gamma}\text{sign}(\omega_n)\xi^2_{{h}}}{2}{f}_{\text{BCS}}
 \end{flalign}
is the value of  $\tilde{f}_t^x$   for zero  SOC, in the F region  below the superconducting electrodes far from the bridge region. The first term in the second line of Eq.(\ref{eq:current_precession}) is  the lowest correction in the SOC which stems from the  precession term in Eq. (\ref{eq:first_order}) which generates the LRTC  $\tilde{f}_t^z$ from rotation of the short-range $\tilde{f}_{t,0}^x$. It is  a positive contribution ($0$-junction) and  as expected depends on the direction of the field.  For $\vartheta=0$, it is only finite if the Rashba SOC is nonzero ({\it cf. } with the numerical results shown in Fig.2). 

In the next order of  the SOC  the contribution to the current is  negative, second term in the second line of Eq. (\ref{eq:current_precession}), and it is due  to the spin relaxation term $\Gamma_{yx}$ in Eq. (\ref{eq:second_order}), that leads to a finite $\tilde{f}_t^y$ component. 
This contribution is only finite if both Rashba and Dresselhaus type of SOC are present. This explains why in the case of a  pure Rashba or Dresselhaus SOC the current does not change sign as a function of $\vartheta$ (see numerical results shown in Fig.2a-b,d-e). 

Thus,   the sign and magnitude  of the critical current is determined by  two competing contributions, namely spin precession and anisotropic spin relaxation\cite{Bergeret2014},  which in turn  depend strongly on the direction of the applied Zeeman field.
For example the contribution  due to spin precession is zero whenever the SU(2) electric field strength in transport direction $\mathcal{F}_{x,0}(\vartheta)=-i\left[\hat{\mathcal{A}}_x,\hat{h}(\vartheta)\right]$ vanishes. This is in accordance with previous theoretical investigations that identified $\mathcal{F}_{k,0}$ as the generator of the LRTC \cite{Bergeret2013,Bergeret2014}.   According to  Eq. (\ref{eq:current_precession}),  $\mathcal{F}_{x,0}(\vartheta)=0$, for $\vartheta_0=\arctan\left(-\frac{\alpha}{\beta}\right)+n\pi$.  For this value of $\vartheta$ the second negative term dominates provided that $\cos 2\vartheta_0\neq 0$, and leads to a change of sign of the critical current, a $0$-$\pi$ transition. 

  For either pure Rashba or pure  Dresselhaus SOC the dependence of  the critical current on $\vartheta$ is simply shifted by $\pi/2$ for the same magnitude of the SOC  parameter. This can be already inferred from Eq.~(\ref{Usadel_basic}), which is symmetric when interchanging $\alpha\leftrightarrow \beta$ and $x\leftrightarrow y$ for the coordinate labels in spin space.


Eq. (\ref{eq:current_precession})  is valid for a symmetric junction, i.e. a junction of type 1 with the same SOC at both electrodes. In the case that the left (L) and right (R) electrodes have different values for the  Rashba and Dresselhaus SOCs $\alpha_{L/R}$ and $\beta_{L/R}$  it is possible to obtain a $0$-$\pi$ transition  solely due to spin precession effects.  Namely, the critical current up to first order in the SOC fields reads
\begin{flalign}\label{eq:opi_precession}
	&j_c=\pi eN_0DT \sum_{\omega_n} \frac{|f_x^b|^2}{\kappa_\omega}e^{-\kappa_\omega L}\times\nonumber\\&(\bar{\alpha}_L \cos\vartheta+\bar{\beta}_L\sin\vartheta)(\bar{\alpha}_R \cos\vartheta+\bar{\beta}_R\sin\vartheta)\; .
\end{flalign}
By inspecting Eq.(\ref{eq:opi_precession}) we see that the current reversal appears every-time the SU(2) electric field strength disappears in the left or right lead $\mathcal{F}_{x,0}^{L/R}=0$, as long as $\alpha_L\beta_R \neq \alpha_R \beta_L$. When all couplings are non vanishing this takes place at the angles $\vartheta_0^{L/R}=\arctan\left(-\frac{\alpha_{L/R}}{\beta_{L/R}}\right)+n\pi$. The interval, where the current is reversed with respect to the symmetric case, is maximized when there is only Rashba SOC in one lead and only Dresselhaus SOC in the other as then  $j_{c}\propto \alpha\beta\sin(2\vartheta)$. 

To summarize this section, for low SOC strength, the long-range  supercurrent  is mainly determined  by the spin precession.   If the S-electrodes are symmetric and only one type of SOC is active, the current can be switched on and off by rotating the exchange field in the $x-y$ plane, but no $0$-$\pi$ transition takes place.  A reversal of the current only appears if both SOC types are finite and originates in a competition of the spin precession- and the spin relaxation effects. A current reversal due to spin precession effects can only be achieved by choosing leads with different SOC parameters.

\begin{figure}
   \begin{center} 
      \includegraphics[width=\columnwidth]{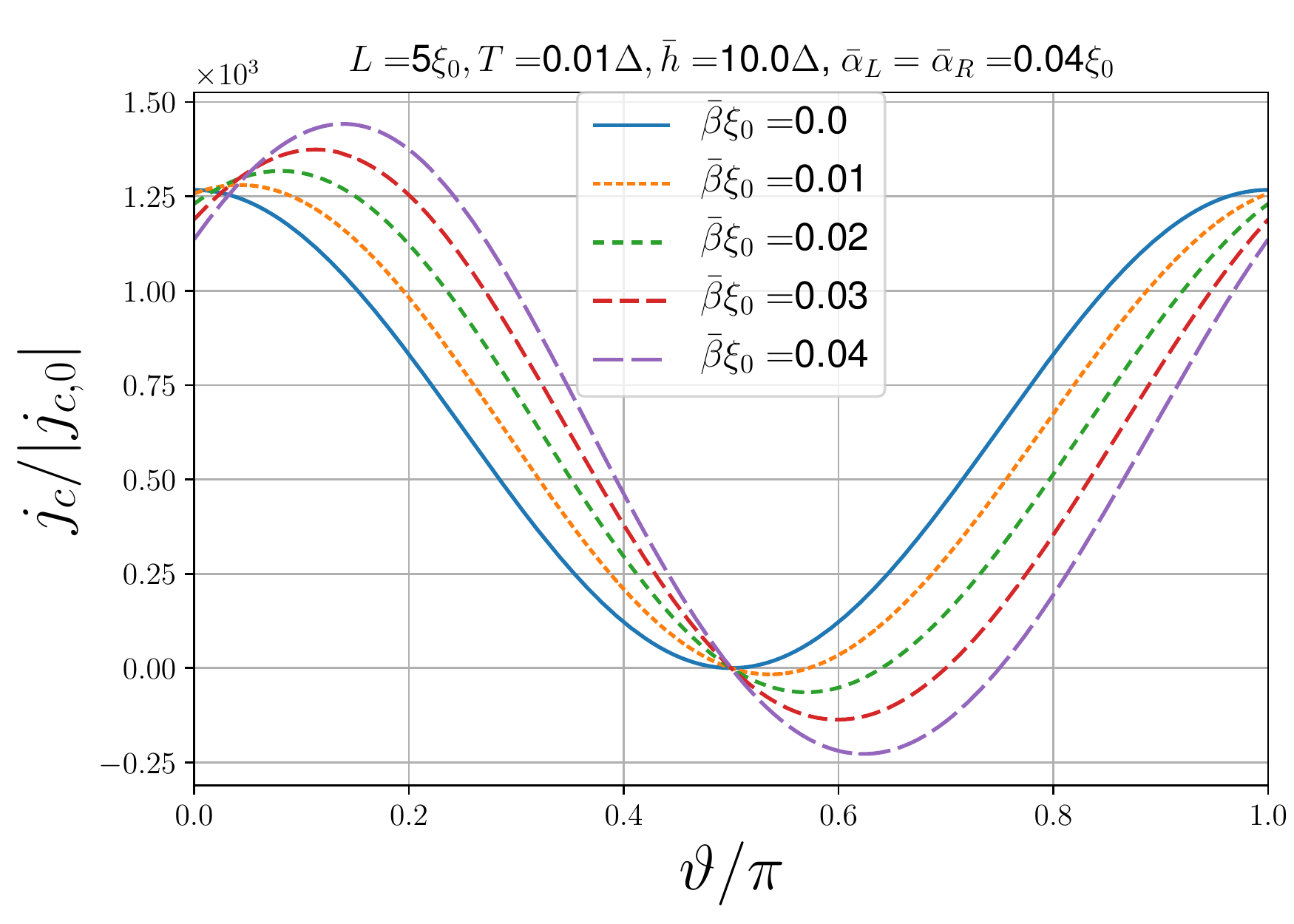}
         \caption{ Numerical results for the critical current as function of the orientation of an in-plane exchange field for an asymmetric junction of type 1. }\label{opi_low}
   \end{center}
\end{figure}

\begin{figure}[ht]
   \begin{center} 
      a)\includegraphics[width=0.9\columnwidth]{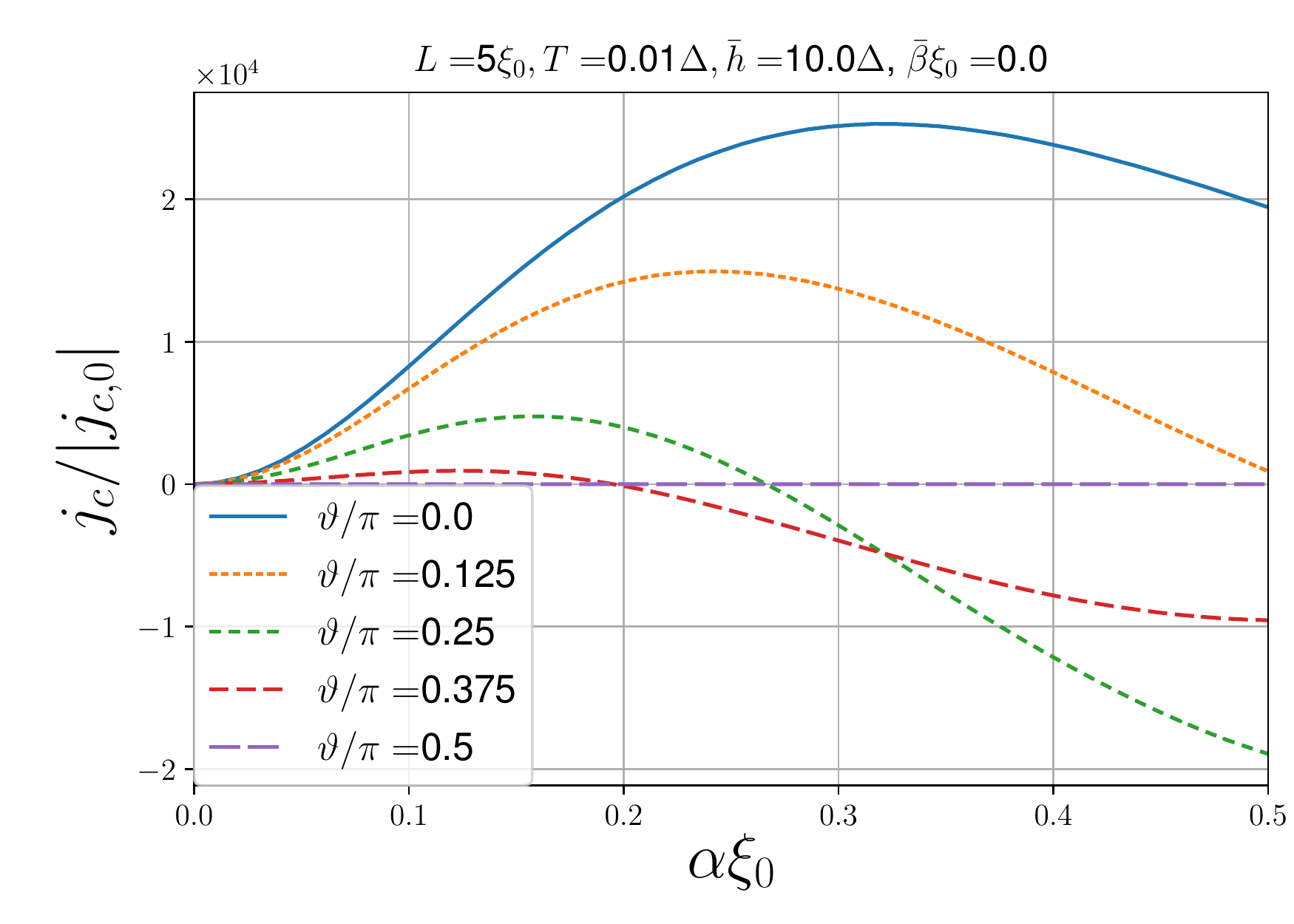}
	        b)\includegraphics[width=0.9\columnwidth]{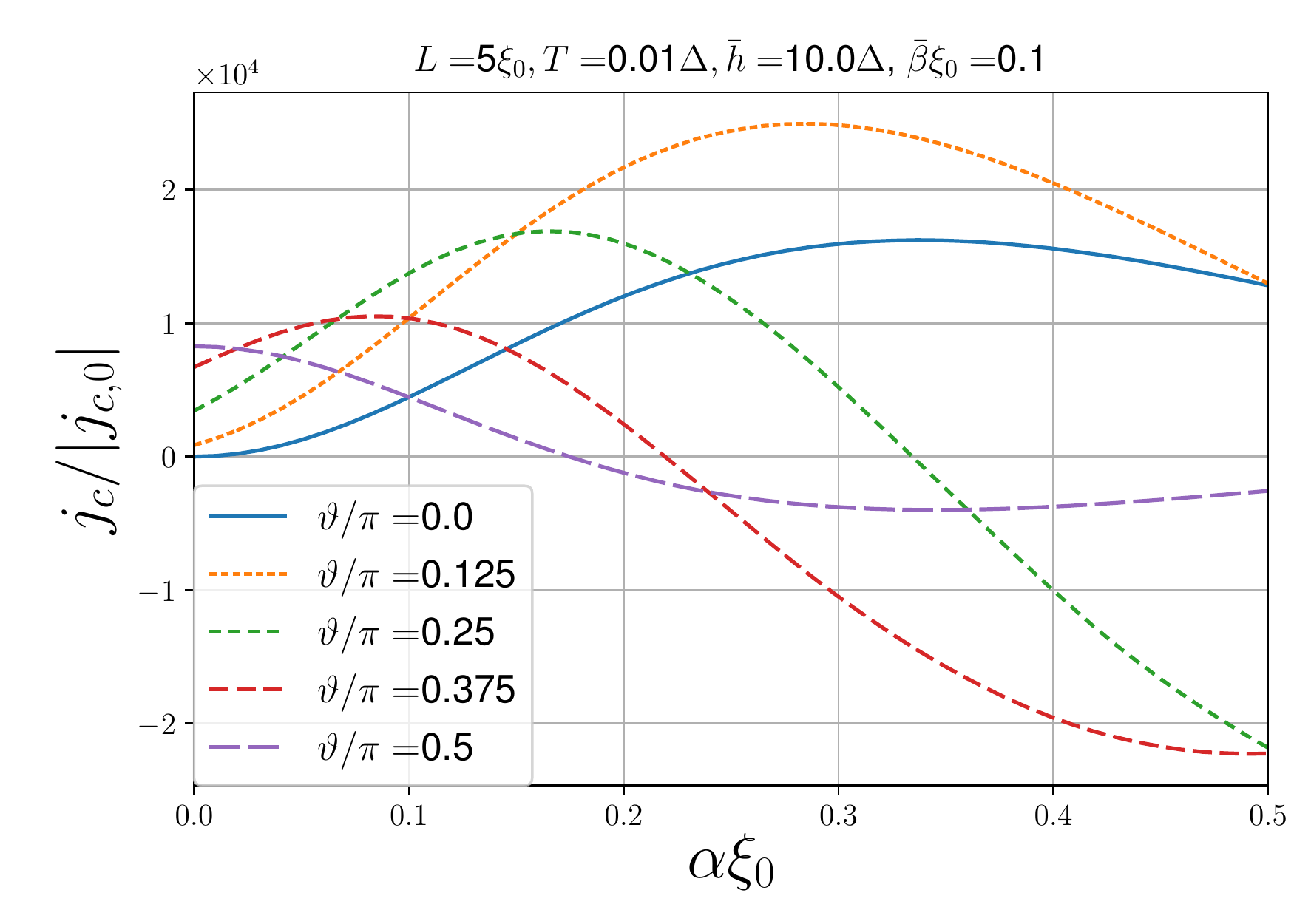}
			      c)\includegraphics[width=0.9\columnwidth]{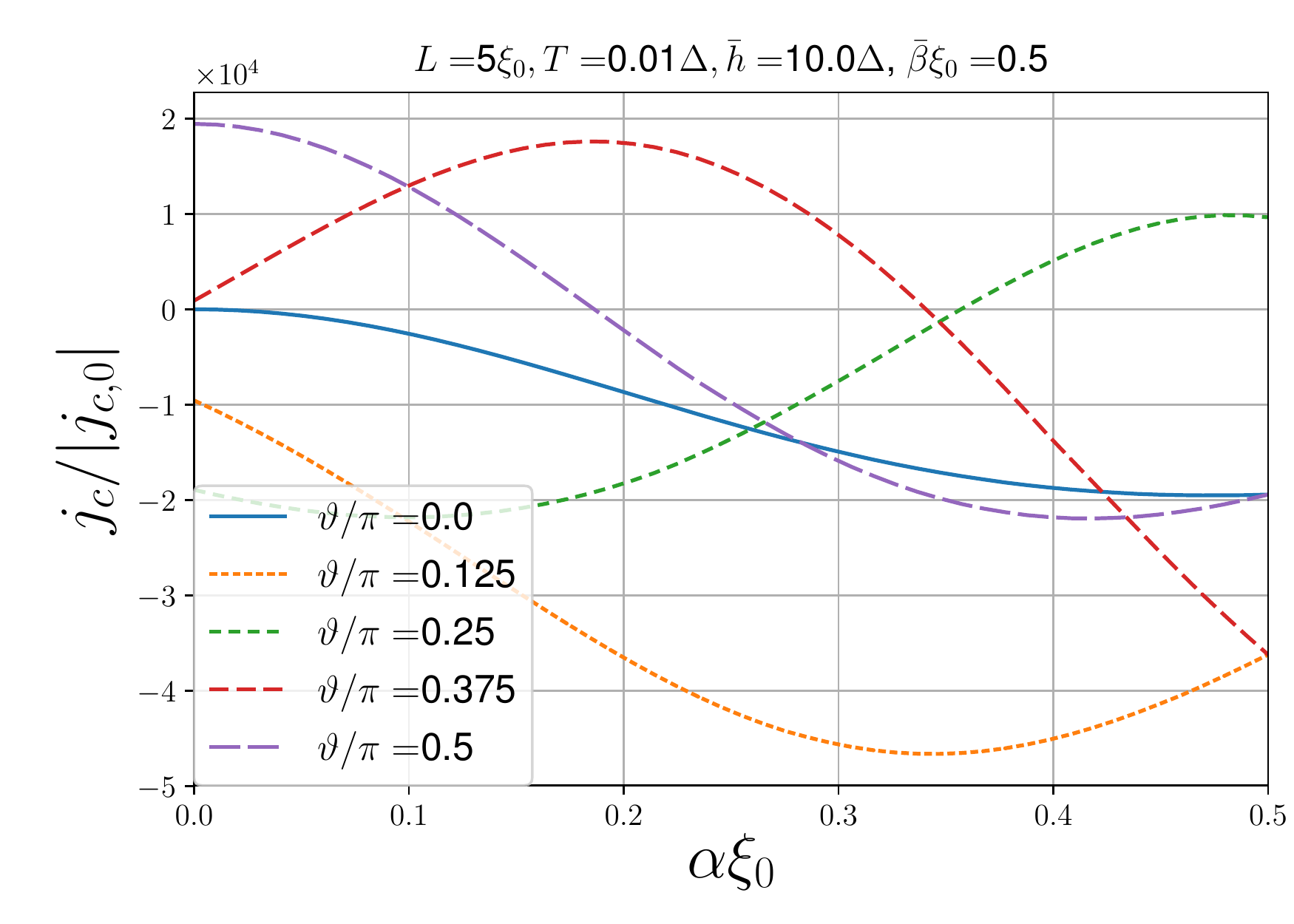}
   \caption{Critical current for a  junction of type 2 as function of the Rashba SOC strength $\alpha$, for different orientations of the exchange field. The  Dresselhaus SOC parameter $\beta$ is increased from  panel a) to panel  c).}\label{J2alpha_varL}
   \end{center}
\end{figure}



\section{Numerical results}\label{section4}
In this section we compute numerically the Josephson current for both types of junctions with finite S-electrodes. 
The total length of the system is  $L_{tot}=2L_S+L$, where $L_S$ is the length of the S-electrode, and is set to $L_{tot}=10L$. 
 The  systems of equations  (\ref{SOC1}), (\ref{eq:ftycomp})  are complemented  by the boundary condition Eq.(\ref{vaccum}) at the outer interfaces: 
\begin{flalign}
	\tilde{\nabla}_x \hat{f}|_{x=\pm {L_{tot}}/{2}}=0.
\end{flalign}
The resulting critical current density for the junction type 1 is shown in Fig.\ref{a_var1} a)-f) and for junction type 2 in Fig.\ref{J2a_var2} a)-f). 
For low SOC strengths and any of the studied SOC types and junction types, the current vanishes when the SU(2) electric field strength vanishes in accordance with previous  theories \cite{Bergeret2013}.  Indeed, the critical current for both setups and small SOC show qualitatively identical behavior (Fig.\ref{a_var1} a)-c), \ref{J2a_var2} a)-c)),  in very good agreement with the analytical result of Eq.(\ref{eq:current_precession}). This implies that at the level of spin-precession effects both junctions behave similarly. As expected the critical current curves for the case of pure Rashba or Dresselhaus SOC are shifted by $\pi/2$ when comparing curves of corresponding SOC strengths.

When increasing the SOC strength for j type 1 junction we observe the competition between the two LRTC generating mechanisms. Comparing the upper and lower panels of Fig.\ref{a_var1} we see that the current changes sign at sufficiently large SOC strengths, only when both Rashba and Dresselhaus SOC are finite, as expected. For the special case when $\alpha=\beta$ and exchange field orientation $\vartheta=\pi/4$ there is no $0-\pi$ transition possible as the spin relaxation contribution to the current vanishes. At $\vartheta=3/4\pi $ spin precession and spin relaxation contributions vanish simultaneously, as can be seen in Fig.\ref{a_var1}f).

By further increase of the SOC the  numerical results shown in Fig.(\ref{a_var1}) f) differ qualitatively from the analytic ones:    there is  a strong increase of the critical current in two negative dips around $\vartheta=\pi/4$. The two negative dips move closer to $\vartheta=\pi/4$ by increasing the  SOC strength. Also there is a flattening of the curve at $\vartheta=3/4$.

 The $0-\pi$ transition  due to spin precession effect in junction type 1 with asymmetric SO interaction obtained analytically in the previous section  is confirmed by the numerics  as shown in Fig.\ref{opi_low}. In particular, the points of current reversal as function of $\vartheta$ are in  agreement with the analytical result,  Eq.(\ref{eq:opi_precession}).

The case of large SOC in  type 2 junctions are shown in  Fig.\ref{J2a_var2} d)-f). We clearly see that  $0-\pi$ transitions  are possible for any choice of SOC. The case when $\alpha,\beta\neq 0$ is qualitatively similar to junction type 1. In contrast, for junction 2, $0-\pi$ transitions are possible for pure Rashba or  Dresselhaus SOC when increasing the SOC strength, as shown in Fig.\ref{J2a_var2} c),d) and Fig.\ref{J2alpha_varL}. 
Our results, regarding the current sign reversal,  are similar   to the results of  Ref.\cite{Arjoranta2016}, where a one dimensional junction with a pure Rashba has been studied. Similarly to the one dimensional case,  our results for two dimensional SOC, show that the direction of the current can be inverted  by tuning the strength of the Rashba SOC,   which can be done by a voltage gate if the bridge region is a semiconductor.  Such a gate  has also been suggested  in Ref.\cite{Liu2014} for creation of a long ranged spin-triplet helix in a ballistic ferromagnetic Josephson junction.

\section{Conclusion}\label{section5}
We present a study of the effects of Rashba and Dresselhaus SO interaction in two types of diffusive lateral Josephson junctions. In the first type the bridge linking the superconducting electrodes  is  a  ferromagnet  and the SOC fields originated from heavy metal interlayers placed between the S leads and the F  bridge. In the second geometry the exchange fields and SOC fields are finite over the whole bridge. In a realistic setup this can be realized by  a a 2D semiconducting bridge  in an external magnetic field.  In both cases we determine  the long-range  triplet Josephson current.  We show how the  magnitude and sign of the supercurrent  can be controlled by varying the direction of the exchange field as well as tuning the strengths of the SOC. Besides their relevance for application as supercurrent valves  such  lateral  junctions can be used as a unequivocal  way of detecting  the long-range triplet component of the condensate in lateral setups.

{\it Note added:} During the preparation of the manuscript we became aware of the very recent work Ref.\cite{Eskilt2019} that studies junction type 1 in great detail. Our work confirms and extends the analytical and numerical results therein as the authors mainly focus on pure Rashba SOC.

\section*{acknowledgements}
BB and FSB acknowledge funding by the Spanish Ministerio de Ciencia, Innovación y Universidades (MICINN) under the project FIS2017-82804-P and by the Transnational Common Laboratory {\it{QuantumChemPhys}}. RB acknowledges funding from the European Union’s Horizon 2020 research and innovation program under the Marie Skłodowska-Curie grant agreement No. 793318.

\newpage
\newpage

\appendix

\section{Basic equations}
After performing the $z$-integration, the resulting system of differential equations for the transformed anomalous Green's function  $\hat{\tilde{f}}=\mathcal{U}\hat{f}\mathcal{U}^\dagger$  for $|x|>L/2$ is:
\begin{flalign}\label{SOC11}
 &D\left[\partial_x^2 \tilde{f}_s\right]-2|\omega_n|\tilde{f}_s-2i\text{sign}(\omega_n) \bar{h} \tilde{f}_t^x=\nonumber\\&-D\bar{\gamma} {f}_\text{BCS}e^{-i\text{sign}(x)\frac{\varphi}{2}}\\
 &D\left[\partial_x^2 \tilde{f}_t^x+2\bar{\mathcal{C}}_x^{xb}\left(\partial_x \tilde{f}_t^b\right)\right]-2|\omega_n|\tilde{f}_t^x- D\bar{\Gamma}^{xb}\tilde{f}_t^b=\nonumber\\&2i\text{sign}(\omega_n)\tilde{f}_s \bar{h}\\
&D\left[\partial_x^2 \tilde{f}_t^y+2\bar{\mathcal{C}}_x^{yb}\left(\partial_x \tilde{f}_t^b\right)\right]-2|\omega_n|\tilde{f}_t^y- D\bar{\Gamma}^{yb}\tilde{f}_t^b=0\\
&D\left[\partial_x^2 \tilde{f}_t^z+2\bar{\mathcal{C}}_x^{zb}\left(\partial_x \tilde{f}_t^b\right)\right]-2|\omega_n|\tilde{f}_t^z- D\bar{\Gamma}^{zb}\tilde{f}_t^b=0.
\end{flalign} 
In the barrier region $|x|<L/2$ we get
\begin{flalign}\label{SOC12}
 &D\partial_x^2 \tilde{f}_s-2|\omega_n|\tilde{f}_s-2{i}\text{sign}(\omega_n)\bar{h} \tilde{f}_t^x=0\\
 &D\partial_x^2 \tilde{f}_t^x-2|\omega_n|\tilde{f}_t^x-2{i}\text{sign}(\omega_n)\bar{h}\tilde{f}_s =0\\
 &D\partial_x^2 \tilde{f}_t^y-2|\omega_n|\tilde{f}_t^y=0\\
 &D\partial_x^2 \tilde{f}_t^z-2|\omega_n|\tilde{f}_t^z=0.
\end{flalign}
The $z$-integration causes a averaging of the couplings as described in the main text. The solution of this system of equations are continuous and fulfill
 \begin{flalign}\label{SOC13}
  &\partial_x \tilde{f}_s \big|_{x=\pm\frac{L}{2}+0^-}=\partial_x \tilde{f}_s\big|_{x=\pm\frac{L}{2}+0^+}\\
  &\label{SOC15}\partial_x \tilde{f}_t^a\big|_{x=\pm\frac{L}{2}+0^\mp}=\left[\partial_x \tilde{f}_t^a +{\bar{\mathcal{C}}}_x^{ab} \tilde{f}_t^b\right]_{x=\pm \frac{L}{2}+0^\pm},
 \end{flalign}
 at the boundaries between the different regions. The spin precession tensor components ${\mathcal{C}}_k^{ab}$ and DP tensor components ${\Gamma}^{ab}$ in the rotated system are determined from the transformed fields 
  \begin{flalign}
 &\hat{\tilde{\mathcal{A}}}_x=\frac{\hat{\sigma}^x}{2}{\zeta}(\vartheta)-{\eta}(\vartheta)\frac{\hat{\sigma}^y}{2},\\ &\hat{\tilde{\mathcal{A}}}_y=\frac{\hat{\sigma}^x}{2}{\eta}(-\vartheta)-{\zeta}(-\vartheta)\frac{\hat{\sigma}^y}{2},
  \end{flalign}
with ${\eta}(\vartheta)=\bar{\alpha}\cos(\vartheta)+\bar{\beta}\sin(\vartheta) $ and ${\zeta}(\vartheta)=-\bar{\alpha}\sin(\vartheta)+\bar{\beta}\cos(\vartheta) $.
 The equations (\ref{SOC11})-(\ref{SOC15}) fully determine the junction system within the limits of the approximations mentioned in the main text. 
 
 \subsection{Zeroth order correction}
 As in the main text we consider the junction type 1 assuming semi-infinite leads. Solving the above system of equations for vanishing SOC gives the following zeroth order solution for the function $\tilde{f}_t^x$, 
 \begin{flalign}\label{eq:ftx_source}
       &\tilde{f}_{t,0}^x
  =\nonumber\\
  &\begin{cases}
  \frac{A_1^L}{\lambda^+}e^{\lambda^+x}
  +
 \frac{ A_2^L}{\lambda^-}e^{\lambda^-x} 
 +
       f_{x}^b e^{-i\frac{\varphi}{2}},x<-\frac{L}{2}\\
	    \frac{B_1}{\lambda^+}e^{\lambda^+x}
	    -
	    \frac{B_2}{\lambda^+}e^{-\lambda^+x}
	   +
	    \frac{B_3}{\lambda^-}e^{\lambda^-x}
	    -
	    \frac{B_4}{\lambda^-}e^{-\lambda^-x}\\
  -\frac{A_1^R}{\lambda^+}e^{-\lambda^+x}
  -\frac{A_2^R}{\lambda^-}e^{-\lambda^-x}
       +
 f_{x}^b e^{i\frac{\varphi}{2}},x>\frac{L}{2}.
 \end{cases}      
 \end{flalign}
 where $\lambda^\pm=\sqrt{\frac{2|\omega_n|}{D}\pm i\frac{2\text{sgn}(\omega_n)\bar{h}}{D}}$,
 \begin{flalign}
  A_{1/2}^L&=\mp \lambda^\pm \frac{f_{s}^b\pm f_{x}^b}{2} \sinh \left(\frac{L\lambda^\pm -i\varphi}{2}\right)\\
  B_{1/2}&=\pm\frac{\lambda^+}{4}\left(f_s^b+f_x^b\right)\text{exp}\left(\frac{-L\lambda^+\pm i\varphi}{2}\right)\\
  B_{3/4}&=\mp\frac{\lambda^-}{4}\left(f_s^b-f_x^b\right)\text{exp}\left(\frac{-L\lambda^-\pm i\varphi}{2}\right)\\
  A_{1/2}^R&=\pm \lambda^\pm \frac{f_{s}^b\pm f_{x}^b}{2} \sinh \left(\frac{L\lambda^\pm +i\varphi}{2}\right) 
 \end{flalign}
 and the bulk solutions for  the singlet and triplet $x$ component
 \begin{flalign}
  &f_{s}^b=D{\gamma}\frac{{f}_{\text{BCS}}}{2}\frac{|\omega_n|}{|\omega_n|^2+{h}^2}\approx \frac{{\gamma}|\omega_n|\xi^2_{\tilde{h}}}{2\bar{h}}{f}_{\text{BCS}}\\
   &f_{x}^b=-iD{\gamma}\frac{f_{\text{BCS}}}{2}\frac{\text{sign}(\omega_n){\bar{h}}}{|\omega_n|^2+\bar{h}^2}
   \approx -i\frac{{\gamma}\text{sign}(\omega_n)\xi^2_{\bar{h}}}{2}{f}_{\text{BCS}}.
 \end{flalign}
 with $\xi_{\bar{h}} =\sqrt{D/\bar{h}}$.

 \subsection{First order correction}

 The Ansatz for the solution of Eq.(\ref{eq:first_order}) reads
 \begin{flalign}
 &\tilde{f}_{t,1}^z(x)=\nonumber\label{eq:first_order_sol}\\
 &\begin{cases}
  K_1 e^{\kappa_\omega x}+Z_1^L e^{\lambda^+ x}+Z_2^L e^{\lambda^- x}, &x<-\frac{L}{2}\\
  K_2 e^{\kappa_\omega x}+
  K_3 e^{-\kappa_\omega x}, &|x|<\frac{L}{2}\\
  K_4 e^{-\kappa_\omega x}+Z_1^R e^{-\lambda^+ x}+Z_2^R e^{-\lambda^- x}, &x>\frac{L}{2}
 \end{cases}
 \end{flalign}
 where
 \begin{flalign} 	&Z_1^L=-\frac{{\eta}(\vartheta)A_1^L}{({\lambda^+}^2-\kappa_\omega^2)},\;\;\;\;Z_2^L=-\frac{{\eta}(\vartheta)A_2^L}{({\lambda^-}^2-\kappa_\omega^2)},\\ 	&Z_1^R=-\frac{{\eta}(\vartheta)A_1^R}{({\lambda^+}^2-\kappa_\omega^2)},\;\;\;\;Z_2^R=-\frac{{\eta}(\vartheta)A_2^R}{({\lambda^-}^2-\kappa_\omega^2)},
 \end{flalign}
Keeping only leading order terms when $\bar{h}\gg T,\text{max}\left\{ \bar{\Gamma}^{ab} \right\}$ and  assuming $L \gg \xi_{\bar{h}}$ we find
 \begin{flalign}
 \left(
  \begin{array}{c}
  K_1\\
  K_2\\
  K_3\\
  K_4
  \end{array}
  \right)\approx\frac{f_x^b}{\kappa_\omega}\frac{{\eta}(\vartheta)}{2}\left(
  \begin{array}{c}
  \sinh(\frac{L\kappa_\omega- i \varphi}{2}) \\
  -\frac{1}{2}e^{-\frac{L\kappa_\omega}{2}}e^{\frac{i\varphi}{2}} \\
  \frac{1}{2}e^{-\frac{L\kappa_\omega}{2}}e^{-\frac{i\varphi}{2}} \\
 -\sinh(\frac{L\kappa_\omega+ i \varphi}{2})                        
  \end{array}\right)
 \end{flalign}

\subsection{Second order correction}
The ansatz for the solution of Eq.(\ref{eq:second_order}) reads
\begin{flalign}\label{eq:second_order_sol}
&\tilde{f}_{t,2}^y(x)=\nonumber\\
&\begin{cases}
 (L_1+x Y_1^L) e^{\kappa_\omega x}+Y_2^L e^{\lambda^+ x}+Y_3^L e^{\lambda^- x}+Y_4^L,\;x<-\frac{L}{2}\\
 L_2 e^{\kappa_\omega x}+
 L_3 e^{-\kappa_\omega x},\;|x|<\frac{L}{2}\\
 (L_4+x Y_1^R) e^{-\kappa_\omega x}+Y_2^R e^{-\lambda^+ x}+Y_3^R e^{-\lambda^- x}+Y_4^R,\;x>\frac{L}{2}
\end{cases}
\end{flalign}
with
\begin{flalign}
	&Y_1^{L/R}=\zeta(\vartheta)K_{1/4},\\
	&Y_2^{L/R}=\pm\frac{2{\lambda^+}^2\zeta(\vartheta)Z_1^L+\bar{\Gamma}_{yx}A_1^{L/R}}{\lambda^+({\lambda^+}^2-\kappa_\omega^2)},\\
	&Y_3^{L/R}=\pm\frac{2{\lambda^-}^2\zeta(\vartheta)Z_2^L+\bar{\Gamma}_{yx}A_2^{L/R}}{\lambda^-({\lambda^-}^2-\kappa_\omega^2)},\\
	&Y_4^{L/R}=-\bar{\Gamma}_{yx}\frac{f_x^b e^{\mp i\frac{\varphi}{2}}}{\kappa_\omega^2}.
\end{flalign}
Considering only leading order terms when $\bar{h}\gg T,\text{max}\left\{ \bar{\Gamma}^{ab} \right\}$ consistent with the first order correction and assuming that $L\gg\xi_{\bar{h}}$ gives for the relevant coefficients inside the bride
\begin{flalign}
\left(
 \begin{array}{c}
 L_2\\
 L_3\\
 \end{array}
 \right)=-\frac{1}{2}e^{-\frac{L\kappa}{2}}\left(
 \begin{array}{c}
 Y_4^R \\
 Y_4^L \\                  
 \end{array}\right).
\end{flalign}


\bibliographystyle{new}
\bibliography{library}

\begin{thebibliography}{10}

\bibitem{Bergeret2001a}
F.~S. Bergeret, A.~F. Volkov, and K.~B. Efetov,
\newblock Phys. Rev. Lett. {\bf 86}, 4096 (2001).

\bibitem{Bergeret2005}
F.~S. Bergeret, A.~F. Volkov, and K.~B. Efetov,
\newblock Rev. Mod. Phys. {\bf 77}, 1321 (2005), 0506047.

\bibitem{Eschrig2011}
M.~Eschrig,
\newblock Phys. Today {\bf 64}, 43 (2011), 1509.02242.

\bibitem{Linder2015}
J.~Linder and J.~W. Robinson,
\newblock Nat. Phys. {\bf 11}, 307 (2015).

\bibitem{Bu1977}
L.~Bulaevskii, V.~Kuzii, and A.~Sobyanin,
\newblock JETP Lett. {\bf 25}, 290 (1977).

\bibitem{Panyukov1982}
S.~V. Panyukov, L.~N. Bulaevskil, and A.~I. Buzdin,
\newblock Sov. Phys. JETP {\bf 35}, 178 (1982).

\bibitem{Ryazanov2001}
V.~V. Ryazanov {\em et~al.},
\newblock Phys. Rev. Lett. {\bf 86}, 2427 (2001), 0008364.

\bibitem{Volkov2001}
A.~F. Volkov, K.~B. Efetov, and F.~S. Bergeret,
\newblock Phys. Rev. B {\bf 64}, 1 (2001).

\bibitem{Kontos2002}
T.~Kontos {\em et~al.},
\newblock Physical Review Letters {\bf 89}, 1 (2002).

\bibitem{Bergeret2013}
F.~S. Bergeret and I.~V. Tokatly,
\newblock Phys. Rev. Lett. {\bf 110}, 1 (2013), 1211.3084.

\bibitem{Bergeret2014}
F.~S. Bergeret and I.~V. Tokatly,
\newblock Phys. Rev. B {\bf 89}, 134517 (2014), 1402.1025.

\bibitem{Robinson2011}
J.~Robinson, J.~Witt, and M.~Blamire,
\newblock Science {\bf 329}, 59 (2010).

\bibitem{Anwar2010}
M.~S. Anwar, F.~Czeschka, M.~Hesselberth, M.~Porcu, and J.~Aarts,
\newblock Phys. Rev. B {\bf 82}, 2 (2010).

\bibitem{Wang2014}
X.~L. Wang {\em et~al.},
\newblock Phys. Rev. B {\bf 89}, 3 (2014).

\bibitem{Gingrich2012}
E.~C. Gingrich {\em et~al.},
\newblock Phys. Rev. B {\bf 86}, 1 (2012).

\bibitem{Robinson2012}
J.~W.~A. Robinson, F.~Chiodi, M.~Egilmez, G.~B. Hal{\'{a}}sz, and M.~G.
  Blamire,
\newblock Sci. Rep. {\bf 2}, 699 (2012).

\bibitem{Chiodi2013}
F.~Chiodi {\em et~al.},
\newblock EPL {\bf 101}, 37002 (2013).

\bibitem{Pal2014}
A.~Pal, Z.~Barber, J.~Robinson, and M.~Blamire,
\newblock Nat. Commun. {\bf 5}, 3340 (2014).

\bibitem{Robinson2014}
J.~W.~A. Robinson, N.~Banerjee, and M.~G. Blamire,
\newblock Phys. Rev. B {\bf 89}, 104505 (2014).

\bibitem{Kalcheim2012}
Y.~Kalcheim, O.~Millo, M.~Egilmez, J.~W.~A. Robinson, and M.~G. Blamire,
\newblock Phys. Rev. B {\bf 85}, 1 (2012).

\bibitem{Banerjee2014}
N.~Banerjee {\em et~al.},
\newblock Nat. Commun. {\bf 5}, 3048 (2014).

\bibitem{Khaire2010}
T.~S. Khaire, M.~A. Khasawneh, W.~P. Pratt, and N.~O. Birge,
\newblock Phys. Rev. Lett. {\bf 104}, 137002 (2010).

\bibitem{Satchell2018}
N.~Satchell and N.~O. Birge,
\newblock Phys. Rev. B {\bf 97}, 1 (2018).

\bibitem{Banerjee2018}
N.~Banerjee {\em et~al.},
\newblock Phys. Rev. B {\bf 97}, 1 (2018).

\bibitem{Satchell2019}
N.~Satchell, R.~Loloee, and N.~O. Birge,
\newblock Phys. Rev. B {\bf 99}, 1 (2019), arXiv:1904.08798v2.

\bibitem{Liu2014}
X.~Liu, J.~K. Jain, and C.~X. Liu,
\newblock Phys. Rev. Lett. {\bf 113}, 1 (2014).

\bibitem{Arjoranta2016}
J.~Arjoranta and T.~T. Heikkil{\"{a}},
\newblock Phys. Rev. B {\bf 93}, 1 (2016).

\bibitem{Eskilt2019}
J.~R. Eskilt, M.~Amundsen, N.~Banerjee, and J.~Linder,
\newblock arxiv  (2019), 1906.07725.

\bibitem{Usadel1970}
K.~Usadel,
\newblock Phys. Rev. Lett. {\bf 25}, 507 (1970).

\bibitem{Note1}
Eq. (\ref {Usadel_basic}) is written in the strict diffusive limit and do not
  take into account charge-spin conversion terms which are higher order in the
  momentum relaxation rate\cite {Bergeret2015,Konschelle2015}.

\bibitem{Kupriyanov1988}
M.~Y. Kupriyanov and V.~F. Lukichev,
\newblock Sov. Phys. JETP {\bf 67}, 1163 (1988).

\bibitem{Konschelle2015}
F.~Konschelle, I.~V. Tokatly, and F.~S. Bergeret,
\newblock Phys. Rev. B {\bf 92} (2015), 1506.02977.

\bibitem{Bergeret2015}
F.~S. Bergeret and I.~V. Tokatly,
\newblock Epl {\bf 110} (2015).

\end{thebibliography}

\end{document}